\documentclass[letterpaper,twocolumn,10pt]{article}
\PassOptionsToPackage{dvipsnames}{xcolor}
\usepackage{usenix}
\usepackage[linesnumbered,ruled,vlined]{algorithm2e}
\usepackage{amsmath}
\usepackage{setspace}

\usepackage{ulem,bookmark}
\usepackage{xurl}
\usepackage{color}
\usepackage{tabularray}
\usepackage[utf8]{inputenc}
\usepackage{multirow} 
\usepackage{threeparttable}  
\usepackage{pifont}
\usepackage{bbding}
\usepackage{fontawesome}
\usepackage{tabularx}
\usepackage{cite}
\usepackage{amsmath,amssymb,amsfonts}
\usepackage{graphicx}
\usepackage{amsthm}%
\usepackage{mathrsfs}%
\usepackage{manyfoot}%
\usepackage{algpseudocode}%
\usepackage{listings}%
\usepackage{siunitx}
\usepackage{indentfirst}
\usepackage{caption}
\usepackage{subfigure}
\usepackage{subcaption}
\usepackage{xspace} 
\usepackage{soul}
\usepackage{titlesec}
\usepackage{makecell}
\usepackage{array}
\usepackage{framed}
\usepackage{booktabs}
\usepackage{pdfpages}
\usepackage{microtype}
\microtypecontext{spacing=nonfrench}

\usepackage{enumitem}
\usepackage{footmisc}
\usepackage{tikz}
\usepackage{diagbox}
\usepackage{CJKutf8}

\usepackage{nicefrac}
\usepackage{
  color,
  float,
  epsfig,
  wrapfig,
  graphics,
  graphicx
}
\usepackage[dvipsnames]{xcolor}
\usepackage{textcomp}
\usepackage{amssymb}
\usepackage{setspace}

\usepackage{latexsym,fancyhdr}
\usepackage{enumerate}
\usepackage{algorithm2e}

\usepackage{graphics}
\usepackage{xparse}

\usepackage{csvsimple}
\usepackage{balance}
\usepackage{tcolorbox}
\usepackage[available]{usenixbadges}
\definecolor{boxbg}{HTML}{F7F8FA}
\definecolor{boxframe}{HTML}{D0D5DD}
\definecolor{checkblue}{HTML}{2563EB}
\definecolor{riskorange}{HTML}{D97706}

\begin{document}

\date{}

\title{\Large Understanding Implicit Trust Errors in Core Carrier Networks through Multi-Agent Flaw Discovery and Analysis}

\author{
  Ziyu Lin, Ziting Wang, Xinfeng Li, Wei Dong, XiaoFeng Wang \\
  \textit{Nanyang Technological University} \\
}

\maketitle

\thispagestyle{empty}

\begin{abstract}

Cellular core networks (CNs) are critical infrastructure, yet their internal security model has historically relied on physical isolation: interfaces between core components often operate within an assumed trust zone. As CNs transition to cloud-native deployments, this assumption weakens, expanding the attack surface and enabling external adversaries to reach previously internal interfaces.
From a root-cause analysis of security flaws reported in GitHub issues for open-source CN implementations, we found a recurring pattern of \textit{blind trust} among CN components. Components may omit syntactic validation, fail to enforce semantic invariants, or allocate resources without checking availability. Once internal interfaces become reachable, these weaknesses can lead to severe impacts such as denial of service and session hijacking. We call these vulnerabilities \textit{implicit trust errors} (\textit{iTrue}).

To detect iTrues and understand their security impacts, we designed \textit{iFinder}, an LLM-driven multi-agent system that summarizes known flaws, distills them into detection patterns, and applies them to discover new iTrues in CN implementations. 
To suppress hallucinations produced by large language models (LLMs), we built an innovative strategy that cross-checks both 3GPP specifications and CN code to capture existing protection missed by the agents. Further, we developed a technique that uses LLMs to generate proof-of-concept (PoC) exploits for potential iTrues and iteratively refine the PoCs by automatically executing them against CN implementations and analyzing results. 
Running iFinder on seven prominent open-source CN implementations, we discovered 84 previously unknown vulnerabilities. Among them, 83 have already been confirmed and 81 have been assigned CVEs. 
Importantly, a session-hijacking flaw has been confirmed on real-world commercial 5G core networks. Our findings highlight the pervasiveness of iTrue risks across the cellular core networks and the urgent need for elevated protection within the original trust domains. We plan to make iFinder publicly available to help enhance the security of cellular core networks. 
\end{abstract}

\section{Introduction}

Cellular core networks are critical infrastructure underpinning everyday communication for billions of users.
Vulnerabilities in these systems can cause large-scale service disruptions and compromise user privacy, making their security of paramount importance~\cite{son2025citesting}.
Historically, core network (CN) security has relied on physical isolation: internal interfaces between network functions operate within a trusted zone, often deployed without strong encryption or mutual authentication~\cite{3gpp33210}.
This trust model, however, is increasingly fragile. As operators migrate CNs to cloud-native deployments, the security perimeter shifts from physically isolated private infrastructure to shared cloud environments with complex network topologies~\cite{barrachina2022cloud}.
Prior work has demonstrated that attackers can reach CN interfaces through cloud misconfigurations or by abusing GTP-U tunnels to inject arbitrary traffic~\cite{zhang2025invade}.
These architectural changes expose internal interfaces to external adversaries, invalidating the implicit-trust assumption that has long underpinned core network security.

\vspace{2pt}\noindent\textbf{Vulnerabilities in CN}. Given these emerging attack surfaces, there is an urgent need for systematic approaches to uncover vulnerabilities in core network implementations. Yet, prior work has largely focused on analyzing interactions either between user equipment (UE) and CN components through the Non-Access Stratum (NAS)~\cite{garbelini2022towards, khandker2024astra, sun20255gc, dong2025corecrisis,bennett2024ransacked}, or between base stations and the CN via protocols such as the Next-Generation Application Protocol (NGAP)~\cite{sun20255gc,bennett2024ransacked} and the GPRS Tunnelling Protocol User Plane (GTP-U)~\cite{bennett2024ransacked}. To our knowledge, no systematic study has ever been done to look into security risks within the CN itself, particularly those arising from inter-component communication that may become externally exposed in cloud deployments, as discussed earlier.

On the other hand, industry has continued to report security-critical flaws affecting the internal CN~\cite{blackduck2022open5gs, nccgroup2021open5gs}. For example, Black Duck's security report~\cite{blackduck2022open5gs} reveals that Open5GS's PFCP implementation lacks proper bounds checking on IE length fields, incurring denial of service (DoS) risks. Additional vulnerabilities have surfaced in GitHub Issues, often involving open-source 5G implementations and simulators such as Open5GS, free5GC, and OpenAirInterface (OAI). Many of these findings, which are frequently uncovered using fuzzing tools such as Defensics~\cite{blackduckfuzzing}, are only isolated samples from a much larger CN security problem space, but they offer a clear glimpse into the kinds of threats these networks face. What remains missing, however, is a systematic study that connects these dots, revealing their underlying relations and, more broadly, clarifying what can go wrong in the CN.

In our research, we take a first step toward understanding CN security by analyzing vulnerabilities reported through GitHub Issues in open-source CN implementations. 
Specifically, we collected all security issues from the GitHub repositories of Open5GS and free5GC, and conducted a systematic root-cause analysis using Claude Code. We designed a Chain-of-Thought (CoT) procedure to guide automated root-cause discovery (e.g., identifying assertion-triggered crashes due to session pool exhaustion) and further clustered these flaws according to their underlying causes. Our analysis revealed a striking similarity across the reported issues. All can be grouped into three broad categories: missing validation of message syntax, missing validation of message semantics, and failure to check resource availability. Collectively, these vulnerabilities reflect a fundamental pattern of \textit{blind trust}: CN components accept internal messages as inherently trustworthy, an assumption rooted in the traditional belief that the CN is walled off from untrusted environments, but one that no longer holds in cloud-native deployments. We call them \textit{implicit trust errors}, or \textit{iTrue}.

\vspace{2pt}\noindent\textbf{Finding iTrues}. To systematically discover iTrues, we developed a multi-agent system, \textit{iFinder}, that summarizes the root causes within each cluster, generalizes these summaries into reusable detection patterns, and applies the patterns to identify additional iTrue flaws in CN implementations. 
Achieving this required addressing a key challenge: distilling cluster-level root causes into detection patterns that are both precise and general. The patterns must capture the essential vulnerability logic while remaining broad enough to cover unseen instances across different protocols and programming languages.
To this end, we engineered prompts to emphasize language-agnostic and protocol-agnostic patterns, explicitly identifying key elements and the expected operations over them. We then use these patterns to guide the agent in scanning CN implementations and extracting iTrue candidates (Section~\ref{sec:preprocessing}). For example, \textit{iFinder} abstracts reports of missing Information Element (IE) checks into a protocol-agnostic pattern: locating downstream uses of an IE, then verifying that its presence and validity are validated earlier in message handling. We pass this pattern to the agent to drive LLM-based backward analysis and flag potential iTrues.

Moreover, naively applying these patterns to CN implementations can yield many false positives, largely due to hallucinations and incomplete reasoning when the underlying LLM lacks sufficient context. In particular, we observed that when only a local code fragment is provided, the agent may flag it as an iTrue even though the required validation is performed elsewhere (e.g., an earlier stage of the processing pipeline) outside the snippet.
At a deeper level, the problem is that protocols typically comprise multiple procedures, each spanning several states. Within a given procedure, security-critical checks may be performed in earlier states rather than in the flagged snippet, so validating an iTrue often requires analyzing the preceding states instead of a single snippet.
However, identifying the code for those preceding states is itself challenging. Feeding the entire codebase to an LLM to ``search'' for relevant checks is often both inefficient and unreliable in large, stateful CN implementations.
To mitigate this issue, we developed a novel \textit{code-specification cross-checking} technique. 
The key idea is to map an iTrue candidate to the protocol procedure it implements, as defined in the relevant specifications, and then map the entire procedure back to the codebase to recover the full execution context. This expanded, procedure-level view enables the agent to determine whether the necessary validation and resource checks are actually enforced, substantially reducing false positives when assessing whether a finding is indeed an iTrue.

The final step of \textit{iFinder} is to generate proof-of-concept (PoC) exploits to confirm the presence of an iTrue flaw. To this end, the agent follows an iterative test-and-refine loop: it deploys a simulator-based testbed to execute PoCs generated by the agent, analyzes execution logs to diagnose failures and extract actionable feedback, and then uses this feedback to revise the PoC. This process repeats until the vulnerability is successfully exploited or a testing budget is exhausted.

Running iFinder on seven prominent open-source CN implementations, we uncovered 84 previously unknown vulnerabilities in the Packet Forwarding Control Protocol (PFCP) and GPRS Tunnelling Protocol Control Plane (GTP-C) protocols. 
Among them, 83 have been confirmed by developers and 81 have been assigned CVEs.
These vulnerabilities have significant security implications, including DoS and session hijacking. 
Moreover, we evaluated these vulnerabilities on two real-world commercial 5G core networks using PoCs derived from open-source platforms, identifying two DoS vulnerabilities and one session-hijacking vulnerability. Both DoS vulnerabilities affect the same commercial 5G core network, and we have received a CVE assignment for one of them (CVE-2026-8232).
The session-hijacking flaw was found in both commercial 5G core networks: one vendor has since fixed the issue, with a CVE assigned (CVE-2026-8233), while the other is still in the remediation process. At a high level, this flaw allows an attacker to inject a PFCP Session Modification Request, causing the User Plane Function (UPF) to forward the victim UE’s uplink traffic to the attacker. We include this vulnerability as a case study in Section~\ref{sec: case studies}.

\noindent \textbf{Contributions.} We list our following contributions below:

\vspace{2pt}\noindent$\bullet$ \textit{New findings}. We present the first systematic root-cause analysis of vulnerabilities reported in open-source CN implementations, showing that they largely arise from \textit{implicit trust} among internal CN components.

\vspace{2pt}\noindent$\bullet$ \textit{New intelligent techniques}. We developed \textit{iFinder}, the first multi-agent system for automated vulnerability discovery in carrier core networks. Our techniques address key technical barriers, and an ablation study demonstrates their effectiveness in reducing hallucinations.

\vspace{2pt}\noindent$\bullet$ \textit{Implementation and evaluation}. We implemented \textit{iFinder} and evaluated it on seven widely used open-source CN implementations, uncovering \textbf{84} previously unknown vulnerabilities and \textbf{81} assigned CVEs, including a session-hijacking flaw confirmed on real-world commercial 5G core networks.

\section{Background}
\label{sec: background}

\vspace{1mm}
\subsection{Cellular Network Overview}

A cellular network consists of user equipment (UE), a radio access network (RAN), and a core network (CN) that implements control-plane and user-plane functions. The RAN is composed of eNodeBs (eNBs) in 4G networks~\cite{3gpp23002} and gNodeBs (gNBs) in 5G networks~\cite{3gpp23501}. Cellular control-plane procedures are typically initiated by the UE and transmitted through the RAN before entering the core network as protocol messages. In 4G systems, this signaling traverses the S1 interface, while in 5G systems it traverses the N2 interface~\cite{3gpp23501,3gpp38413}. These access-facing interfaces form the primary boundary between the external RAN and the internal CN. Although this boundary is often treated as trusted in operational deployments, the signaling originates from UE interactions and directly influences CN state. As a result, the RAN-CN boundary has been a recurring focus in security analyses of cellular systems \cite{hussain2018lteinspector, bennett2024ransacked, khandker2024astra}.

\begin{figure}[t]
  \centering
  \includegraphics[width=\linewidth]{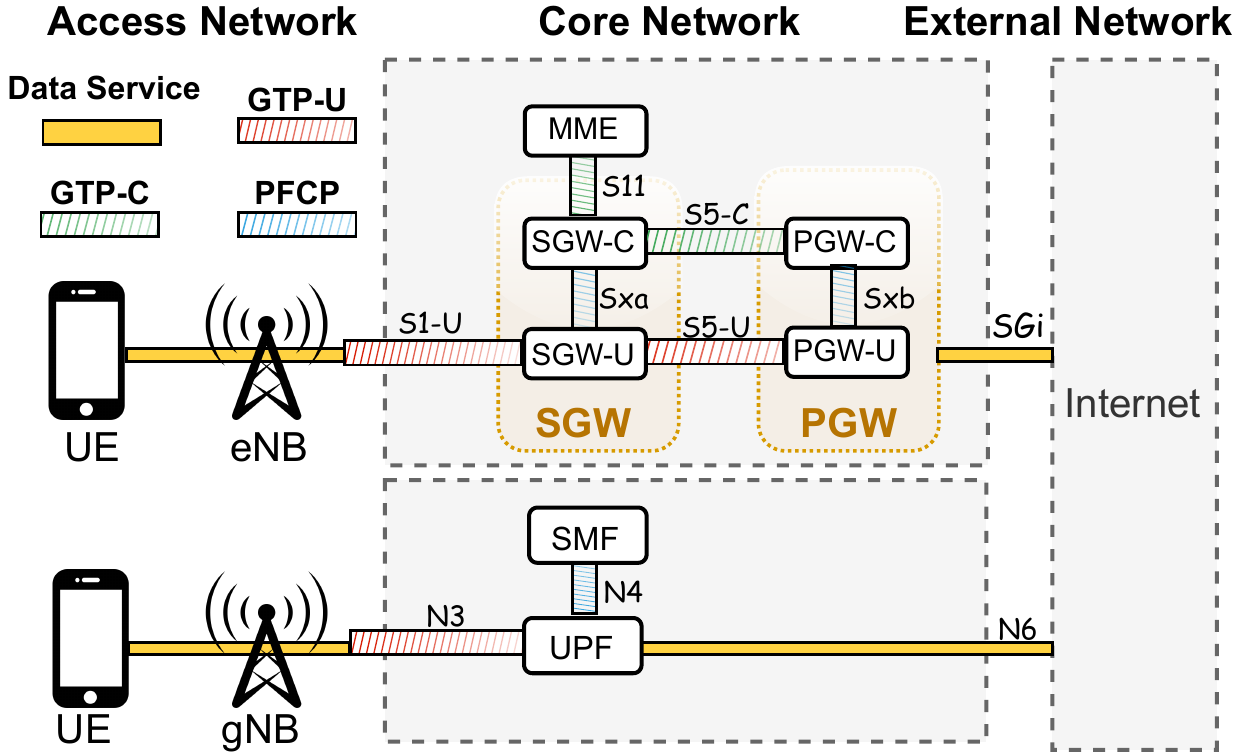}
  \caption{Cellular network architectures from 4G to 5G.}
  \label{fig:cn_arch}
\end{figure}

As shown in Figure~\ref{fig:cn_arch}, the architecture of the core network has evolved significantly from 4G to 5G, with important implications for trust and attack surface. In the 4G Evolved Packet Core, control-plane logic is largely centralized in the Mobility Management Entity (MME)~\cite{3gpp29274}, while the Serving Gateway (SGW) and Packet Data Network Gateway (PGW) anchor user-plane traffic and connectivity to the Internet. Although control and user planes are conceptually separated, their protocol logic and session state are closely coupled across a small set of components. The 5G Core further decomposes these roles into the Access and Mobility Management Function (AMF), the Session Management Function (SMF), and the User Plane Function (UPF), with AMF handling access and mobility control, SMF managing session state, and UPF performing packet forwarding~\cite{3gpp23501}. These functions interact through a larger set of internal interfaces, including N2, N4, and N9. While this decomposition improves flexibility and scalability, it also relies more heavily on implicit trust between internal network functions~\cite{akon2023formal}. In practice, internal interfaces are often deployed without strong cryptographic protection to prioritize performance~\cite{3gpp33210,donegan2011ipsec}, which increases the impact of protocol misuse and configuration errors inside the core network as the architectural gap between the cellular core and the Internet continues to close~\cite{zhang2025invade,akon2023formal}.

\subsection{Core Network Protocol}

Core network behavior is governed by a small set of protocols that manage session establishment, state propagation, and packet forwarding. Among these, Packet Forwarding Control Protocol (PFCP) and GPRS Tunnelling Protocol (GTP) are central to how control-plane decisions are translated into user-plane behavior~\cite{3gpp29274,etsi129244}. PFCP governs control-plane interaction between the SMF and the UPF, while GTP supports both control signaling through GTP-C and user-plane tunnelling through GTP-U across 4G and 5G cores~\cite{3gpp29281,3gpp29274}. These protocols do not operate as isolated message exchanges; instead, they establish long-lived session contexts shared across multiple network functions. System behavior therefore depends on protocol execution paths that span multiple messages and handlers rather than on individual packets~\cite{zhang2025invade,akon2023formal}.

PFCP operates on the N4 interface and is responsible for installing forwarding rules in the user plane. A PFCP association is first established to exchange node identifiers, after which session-specific state is created through PFCP session establishment procedures~\cite{etsi129244}. Each session is identified by a Session Endpoint Identifier (SEID) that is referenced in subsequent operations such as modification and deletion. Because PFCP directly links control-plane signaling to concrete forwarding resources, message-level validity alone does not guarantee safe system behavior. Errors in session state or control logic can immediately lead to traffic hijacking or denial-of-service~\cite{amponis2022threatening,zhang2025invade}. GTP underpins user-plane communication and separates control signaling from data forwarding through GTP-C and GTP-U~\cite{3gpp29060,3gpp29281}. GTP-U remains the primary mechanism for tunnelling user traffic across the core network~\cite{3gpp29281,zhang2025invade}. GTP relies on Tunnel Endpoint Identifiers (TEIDs) to associate packets with tunnels, with each tunnel being unidirectional and identified by a receiver-assigned identifier. Bidirectional communication therefore requires coordinated tunnel setup through a sequence of control-plane messages~\cite{3gpp29274,3gpp29281}. GTP messages also carry many Information Elements (IEs), such as access point settings and quality-of-service (QoS) parameters, which must be parsed and validated consistently across different handlers. 

\subsection{Agents and LLMs}
The complexity of cellular core networks has motivated the use of automated analysis techniques such as fuzzing and static analysis. More recently, large language models have been explored as an additional tool for software analysis~\cite{sheng2025llms}. However, these approaches face fundamental limitations when applied to core network software due to a mismatch between their analysis models and the stateful, procedural nature of cellular protocols. Coverage-guided fuzzers are effective at exploring input-handling logic but struggle to reach protocol states that require multiple rounds of valid message exchanges and coordinated state updates across components~\cite{klees2018evaluating, pham2020aflnet, natella2022stateafl}. Static analysis tools face similar challenges, as control logic is often distributed across asynchronous handlers and shared session contexts, which makes precise tracking of execution paths difficult and leads to false positives or missed flows \cite{hussain2018lteinspector,kim2019touching}.

These limitations motivate the use of agent-based workflows as an orchestration mechanism rather than as a replacement for existing tools. In this setting, an agent coordinates protocol-aware reasoning, staged analysis, and runtime feedback to align analysis steps with protocol procedures and session evolution. By incorporating feedback from concrete executions and analyzing logs to classify triggered impacts~\cite{deng2024pentestgpt,ullah2025cve}, agent-based workflows can bridge abstraction gaps that limit traditional techniques. This approach, leveraging Chain-of-Thought reasoning~\cite{wei2022chain} and specification-grounded cross-checking~\cite{3gpp23501,etsi129244}, is particularly relevant for cellular core networks, where security-relevant behavior emerges from the interaction between complex protocol logic and operational trust assumptions~\cite{deng2024pentestgpt,yao2022react}.

\subsection{Threat Model}
\label{sec: threat model}

\begin{figure}[t]
  \centering
  \includegraphics[width=\linewidth]{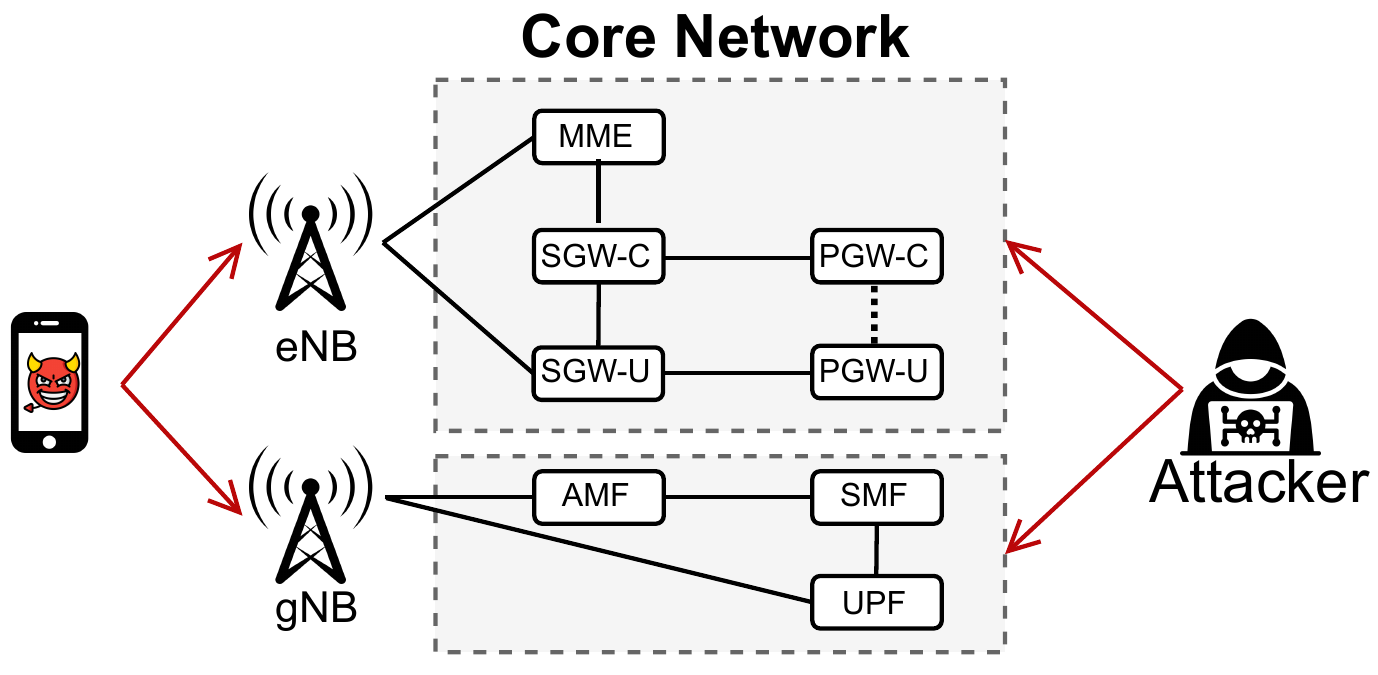}
  \caption{Threat model. Attackers are assumed to have access to internal core network interfaces and can send protocol messages that violate implicit trust assumptions, without compromising cryptographic protections.}
  \label{fig:threat model}
\end{figure}

We assume an adversary can obtain the IP address of core-network components. Such information may be derived from public documentation, passive enumeration, or active scanning~\cite{zhang2025invade}. As shown in Figure~\ref{fig:threat model}, we consider two classes of attackers, distinguished by their network vantage point and capabilities. The two attackers aim to disrupt cellular network services, which can result in DoS and session hijacking.

\vspace{2pt}\noindent$\bullet$ \textit{Internet attacker}. 
The attacker is located outside the cellular core network (e.g., on the public Internet) and does not control any core-network component. However, the attacker can craft and send arbitrary PFCP or GTP-C messages. In cloud-native deployments, misconfigurations may inadvertently expose internal interfaces to external networks. In this case, an attacker can send malformed PFCP or GTP-C messages directly to core components. The presence of such risks in real-world cloud environments has also been reported by prior research~\cite{zhang2025invade}.

\vspace{2pt}\noindent$\bullet$ \textit{Malicious UE}. 
The attacker acts as a malicious UE using a commercial smartphone with a valid Universal Subscriber Identity Module (USIM). After completing the standard registration procedure, the UE obtains a legitimate IP address from the network. The attacker then injects carefully crafted payloads into the uplink data stream. The gNB encapsulates this traffic into GTP-U and forwards it through the UPF. These payloads may carry PFCP or GTP-C messages. 
By exploiting protocol tunnelling and network boundary bridging, the attacker smuggles crafted PFCP or GTP-C messages inside GTP-U messages so that, absent strict boundary enforcement, they cross the boundary and are delivered to and parsed by core-network components~\cite{shaik2025uncovering, mitre_fight_fgt1599_505}.
We further confirmed that both protocol tunnelling and network boundary bridging are feasible against five of the seven evaluated open-source CN implementations.

\section{Understanding Weaknesses in CNs}

Under our threat model, attackers can reach core network components from outside the trusted boundary, as shown in prior work~\cite{zhang2025invade}, invalidating the traditional isolation assumption that internal components only receive well-formed, benign messages. As a result, we discovered that implicit-trust-driven minimal validation becomes an exploitable attack surface. We made this observation through an empirical analysis of security issues reported in CN implementations’ GitHub repositories. Specifically, we extracted the root cause of each issue, clustered issues with similar causes, and distilled recurring vulnerability patterns. We then utilized these patterns to guide an LLM in identifying previously unseen iTrue vulnerabilities across the codebases.

\noindent\textbf{GitHub security issues in CN}. 
To study the relations among CN-related known vulnerabilities, we curated a dataset of previously reported GitHub security issues in CN. Specifically, we gathered 22 security issues from two widely used open-source CN implementations, Open5GS and free5GC. These reports follow a largely consistent template, typically including a title and bug description, the affected version and components, steps to reproduce, and supporting artifacts such as logs, expected and observed behavior.
These fields provide context to localize the faulty checks, enabling systematic root-cause analysis across issues.

\vspace{2pt}\noindent\textbf{Root-cause analysis and clustering}. Over the reported security issues, we performed a root-cause analysis using Claude Code to extract three elements from each issue: the message(s) that introduce security risk (the \textit{trigger}), the program location in CN implementations where the risk manifests (the \textit{failure point}), and the underlying vulnerability (the \textit{flaw}). We then summarized these elements and their relations into a \textit{root-cause statement}. For example, in \texttt{ogs\_pfcp\_handle\_create\_urr()}, an out-of-bounds read (failure point) can be triggered by a Session Establishment Request that omits mandatory fields (e.g., Downlink Volume) in the Volume Threshold IE (trigger), exposing a missing mandatory-field check (flaw) and causing a crash.
Although these issues span different messages and code paths, many root-cause statements share the same structure: the implementation consumes untrusted message fields without validating mandatory IEs, lengths, or bounds. For instance, in \texttt{ogs\_pfcp\_extract\_node\_id()}, the code copies Node ID data into a fixed-size buffer using a length taken directly from an unverified IE length field in an Association Setup Request, due to missing IE-length validation. While the specific IE types and messages differ, both this case and the above example stem from the same underlying cause, that is, failure to verify the format and completeness of incoming messages. Guided by these similarities, we leveraged an LLM to infer relations among issues and obtained three distinct clusters.

\vspace{2pt}\noindent\textbf{Analysis and findings}.
Looking into these clusters, we found that CN components often forgo due diligence in validating message format, message semantics, and resource availability, and instead tend to \textit{blindly} act on messages received from internal peers: 

\vspace{2pt}\noindent$\bullet$ \textit{Missing syntactic validation}: 12 issues involve malformed messages being accepted without checking the presence of mandatory fields or enforcing basic format and length constraints. For example, on receiving a Session Establishment Request containing a Volume Threshold IE, the UPF \textit{blindly} proceeds to copy volume fields based on presence flags, without verifying that all required fields are included. As a result, one can craft a truncated Volume Threshold IE to trigger an out-of-bounds read and crash the UPF.

\vspace{2pt}\noindent$\bullet$ \textit{Missing semantic validation}: 8 issues involve messages being accepted without validating semantic constraints or enforcing state-dependent invariants. For example, upon receiving a Session Establishment Request containing an F-TEID value of \texttt{0}, the UPF \textit{blindly} uses the F-TEID to look up the corresponding PDR without checking whether the value falls within a valid range. As a result, an attacker can craft a PFCP message with an out-of-range F-TEID to trigger an assertion failure and crash the UPF.

\vspace{2pt}\noindent$\bullet$ \textit{Missing checking on resource availability}: 2 issues involve risky resource-handling operations where the recipient allocates requested resources without checking availability. For example, upon receiving a PFCP Session Establishment Request, the UPF \textit{blindly} allocates buffers from its internal pool without verifying that sufficient resources remain. As a result, an attacker can repeatedly send PFCP messages to drain the buffer pool, ultimately triggering a segmentation fault and crashing the UPF.

Across these clusters, the common denominator is the \textit{implicit trust} that a CN component places in messages from its internal peers, rooted in the assumption that the CN is well protected and walled off from untrusted environments. This assumption becomes increasingly fragile as CNs move to the cloud: as discussed earlier, new attack surfaces emerge through exposed CN components and loosely protected tunnel interfaces (Section~\ref{sec: threat model}). 
Consequently, we refer to the risks arising from this legacy ``blind faith'' as \textit{implicit trust errors} (\textit{iTrue}).
This observation inspired us to summarize the defining patterns of iTrues to guide the discovery of previously unknown flaws in CN implementations. Finally, although these known issues are largely memory-safety bugs, in part because they were most likely detected by fuzzing, \textit{we show later that the same patterns enable the discovery of more subtle and more severe vulnerabilities, including those that enable session hijacking} (Section~\ref{sec: case studies}).

\section{Finding iTrues with iFinder Agents}

Given the repeated appearance of iTrue flaws, we hypothesize that they are not isolated incidents but recurring weaknesses across core-network (CN) implementations. We therefore aim to distill the distinctive characteristics of known iTrue instances into a feature set that can guide an LLM to detect iTrues in diverse CN codebases and programming languages. However, we found that naively applying LLM inference is unreliable: it frequently produces hallucinated findings, leading to low precision and inconsistent results. Specifically, in our preliminary evaluation, direct pattern-guided LLM inference yielded only 8 true positives but 56 false positives and 14 false negatives (Precision 28.205\%, Recall 36.364\%, F1 31.769\%), indicating severe context-induced hallucinations. Most false positives arise when required validations are implemented earlier in the same protocol procedure, but are invisible in the local snippet. In contrast, many false negatives are due to the large search space. Consequently, when scanning an entire codebase, the LLM fails to focus on the right protocol-handling paths and misses the vulnerable snippet. 

To address these challenges, we developed a multi-agent system, \textit{iFinder}, which integrates LLM-based backward analysis, code-specification cross-checking, and proof-of-concept (PoC) exploit generation and refinement to improve detection quality. These techniques and the design of the whole system are elaborated in the rest of the section.

\subsection{Overview}

\noindent\textbf{Idea}. To reduce false negatives, we focus on likely failure points, such as operations that depend on message content (e.g., reading or copying based on a length field carried in a message), and then trace back to the \textit{trigger}, such as the message-handling logic, to determine whether the expected validation is present. We also observe that a major source of hallucinations is missing protocol context, which leads the LLM to fixate on irrelevant code segments and report vulnerabilities that are infeasible along real execution paths. This issue is especially common in carrier protocols, where a single procedure spans multiple states and security-critical checks are often performed in earlier states rather than in the flagged snippet (see Figure~\ref{fig:crosscheck-case}). To mitigate this, we propose \textit{code-specification cross-checking}: grounding each vulnerability candidate in the corresponding procedure defined by 3GPP specifications, recovering the prerequisite messages, and then mapping these messages to their handlers in the codebase by searching, so the agent can reason about whether the candidate is still feasible. This procedure-level context substantially reduces context-induced false positives without requiring the LLM to reason over the entire codebase.

To further reduce false positives, we use the LLM to generate proof-of-concept (PoC) exploits for identified iTrue candidates. A key challenge is that LLM-generated PoCs may fail to trigger a flaw due to implicit runtime constraints that are not evident from static code alone. Our idea is \textit{feedback-aware refinement}, which frames exploitation as an iterative process: the agent executes the PoC against a testbed and analyzes runtime logs. If the vulnerability is triggered, the agent confirms the finding; otherwise, it diagnoses the failure and extracts actionable feedback to refine the PoC accordingly. This test-and-refine loop grounds PoC generation in observed execution behavior, reducing exploitation-time hallucinations and improving the success rate of vulnerability confirmation.

\vspace{2pt}\noindent\textbf{Design}. Built upon these techniques, \textit{iFinder} adopts a multi-agent design to perform four key tasks: preprocessing, discovery, vetting, and exploitation. Every agent in \textit{iFinder} is an LLM session (realized with the Claude Agent SDK) driven by a task prompt and tool calls (\texttt{Grep}, \texttt{Glob}, \texttt{Read}, and build/execute), through which all reasoning is carried out by the LLM.
The preprocessing stage supports the other tasks through two one-time operations (Section~\ref{sec:preprocessing}). First, it extracts protocol procedures and message schemas from 3GPP specifications~\cite{etsi129244}. Second, it summarizes detection patterns from cluster-level root causes; these patterns are designed to be both general, applicable across protocols and programming languages, and accurate.
Guided by these patterns, a Discovery Agent detects iTrue candidates via LLM-based backward analysis. However, this approach can yield a high false-positive rate when the required checks are implemented earlier in the same procedure, as discussed above. To address this challenge, \textit{iFinder} invokes a Vetting Agent to perform code-specification cross-checking (Section~\ref{sec:discovery agent}).
The resulting candidates are further validated by an Exploitation Agent, which automatically generates a PoC from the inferred attack vector and compliant prerequisite messages, executes it against a testbed, and iteratively refines it using real-time runtime logs (Section~\ref{sec: exploitation agent}).

\begin{figure*}[t]
  \centering
  \includegraphics[width=0.75\textwidth]{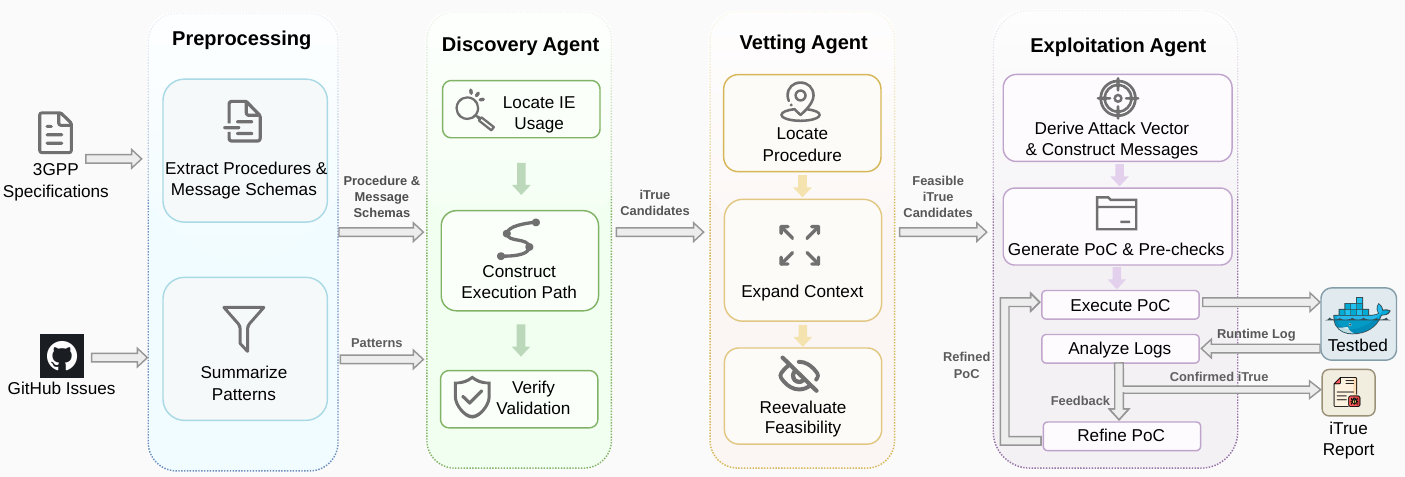}
  \caption{Overview of the iFinder framework. iFinder consists of three agents: a Discovery Agent that identifies candidate implicit trust violations via LLM-based backward analysis, a Vetting Agent that cross-checks code and protocol semantics to filter false positives, and an Exploitation Agent that validates security impact in a controlled testbed.}
  \label{fig:system}
\end{figure*}

These agents communicate via explicit intermediate artifacts, including detection patterns, iTrue candidate reports, feasible-candidate reports, and prerequisite messages, which ensures reliable information flow across agents and supports end-to-end iTrue discovery and confirmation. Figure~\ref{fig:system} illustrates the system workflow. In what follows, we present the designs of these agents and the preprocessing component. 

\subsection{Preprocessing}
\label{sec:preprocessing}
As mentioned earlier, the preprocessing stage aims at (1) extracting procedures and message schemas from 3GPP documents, and (2) summarizing detection patterns distilled from cluster-level root causes. To ensure the reliability of the outcomes, particularly the quality of the patterns, a domain expert reviewed the generated patterns and, when necessary, re-invoked the generation process with additional feedback to improve them.

\vspace{2pt}\noindent\textbf{Summarizing detection patterns}. 
CN implementations are written in heterogeneous languages (e.g., C and Go) and span multiple protocols (e.g., PFCP and GTP-C); thus, a pattern that is both precise and general is necessary to enable more effective and reliable exploration.
To achieve generality, we carefully design prompts so that the generated patterns are expressed in a protocol-agnostic and language-agnostic manner.
Meanwhile, to ensure precision and reduce ambiguity for iTrue discovery, we formulate each pattern as a triple: <element, dangerous operation, missing validation>, where element denotes the involved information element (IE), dangerous operation denotes the security-critical operation performed on the IE, and missing validation denotes the necessary check that is absent before the operation is performed. Figure~\ref{fig:pattern} presents an example that can be instantiated as <mandatory IE, direct dereference, missing presence>. 
Finally, we summarize the six iTrue patterns.

\begin{figure}[t]
\centering
\begin{tcolorbox}[
  colback=boxbg,
  colframe=boxframe,
  coltitle=black,
  title=Absent Field Pattern,
  width=\linewidth,
  boxrule=0.6pt,
  arc=2mm,
  left=1.2mm,right=1.2mm,top=1.0mm,bottom=1.0mm
]
The code directly dereferences the mandatory information element without
\textcolor{checkblue}{checking that the mandatory element is actually present.}
\textcolor{riskorange}{These unchecked accesses can lead to null pointer dereferences, leading to DoS.}
\end{tcolorbox}
\caption{Example of an iTrue detection pattern (\textit{Absent Field}): a mandatory IE is dereferenced without a presence check, enabling a DoS.}
\label{fig:pattern}
\end{figure}

\vspace{2pt}\noindent\textbf{Extracting procedures and message schemas from specifications}. 
iFinder utilizes an instruction prompt to extract procedures and message schemas from 3GPP specifications. The prompt includes an illustrative example (e.g., the Modify Bearer procedure) to explain what constitutes a procedure and reduce ambiguity. Such extracted knowledge helps downstream agents better understand protocol behavior,
enabling them to reason about protocol execution when validating flagged iTrues and generating PoCs.

\subsection{Discovery Agent}
\label{sec:discovery agent}
With the inputs from the preprocessing component, particularly the summarized patterns, the Discovery Agent (DA) searches for iTrue instances across various CN implementations. 
As mentioned earlier, directly prompting an LLM to scan the entire codebase and flag iTrue candidates matching a pattern is impractical: the codebase spans over one million lines of code, far exceeding typical context-window limits and making long-context reasoning both inefficient and prone to hallucinations.
To address this challenge, the DA performs \textit{LLM-based backward analysis} for iTrue discovery. 
The DA is an LLM session driven by a task prompt and carries out this analysis entirely through iterative \texttt{Grep}/\texttt{Glob}/\texttt{Read} tool calls.
Specifically, the procedure has three steps: it first locates a dangerous operation that matches the pattern, then walks backward along caller chains to construct the execution path, and finally checks whether the validations required by the pattern are missing.

\noindent\textbf{Locating risky IE usages}.
The DA first identifies code locations where protocol-controlled Information Elements (IEs) are used in a manner that matches a given pattern. 
As described earlier, each detection pattern is specified to be language-agnostic (e.g., ``memory operations'') to support cross-language applicability. 
The DA begins by reading the pattern description and grounding these abstract operations into concrete, language-specific constructions before detecting them in the target CN implementation. For each detected usage site, the DA then determines which IE field is used as an argument to the dangerous operation.
For example, given a ``Malformed Field'' pattern, the DA first locates \texttt{memcpy}-like memory operations; then it finds that \texttt{flow\_description\_len} from the \texttt{SDFFilter} IE is passed as the size argument to a memory operation at \texttt{types.c:340}.

\noindent\textbf{Constructing execution path}.
After locating a pattern-matched IE usage, the DA constructs the execution path by iteratively searching for and inspecting the callers along the backward chain, so that it can later check whether the required validations are present. Specifically, the DA iteratively expands the call chain discovered from a CN implementation by using \texttt{Grep} to identify candidate callers, \texttt{Glob} to locate relevant source files, and \texttt{Read} to inspect each caller's code, repeating these operations until it reaches the protocol message handler\footnote{We found that the agent's standard approach of constructing call graphs via Abstract Syntax Trees (ASTs) is less effective than a simple \texttt{Grep}-based method, possibly because the LLM is better optimized for such document-search commands.}. The resulting call chain spans from the protocol message handler, through intermediate parsing functions, to the operations matching the pattern.

\noindent\textbf{Checking the existence of validation}.
With the execution path constructed, and guided by the pattern specification, the DA checks whether the necessary validations for the target IE appear before the risky use. 
Depending on the pattern, the DA may perform three kinds of validation: (1) protocol-syntax validation (e.g., the presence of mandatory field and basic format/length constraints before accessing IE fields), (2) protocol-semantics validation (e.g., range checks or state-dependent invariants on IE values), and (3) resource-availability validation (e.g., capacity checks or access control before allocating requested resources). 
The DA reads the functions in the execution path to look for these checks; if it does not find the required validation before the dangerous operation, it flags the site as an iTrue candidate.
\subsection{Vetting Agent}
\label{sec: vetting agent}

As mentioned earlier, the vulnerability candidates produced by the DA include many false positives, because security checks may occur in earlier states rather than in the flagged snippet. 
Addressing this challenge is the Vetting Agent (VA), which filters candidates via code-specification cross-checking: a three-step approach that leverages 3GPP specifications to expand the analysis context and reason about each candidate's feasibility. The approach includes locating an iTrue candidate's message in the procedure of 3GPP specifications, then mapping the prerequisite messages to code by searching, and reevaluating iTrue candidates under the expanded code context.

\vspace{2pt}\noindent\textbf{Locating an iTrue candidate in specifications}. 
The first step is to locate the procedure that carries an iTrue candidate in the specifications. For this purpose, the VA first identifies, among the procedures extracted during the preprocessing stage, the one that contains the message involved in the candidate (as identified by the DA). It then enumerates all other messages in the same procedure that are exchanged prior to the triggering message; these messages serve as ``anchors'' to help determine, later, the code segment that implements the procedure. 
For example, consider an iTrue candidate: the assertion \texttt{ogs\_assert\_if\_reached()} at \texttt{s11-handler.c:608}, which is triggered along the \texttt{ModifyBearerRequest} handling path when \texttt{session\_type} takes an unexpected value. Using the triggering message \texttt{ModifyBearerRequest}, the VA maps the candidate to the Modify Bearer procedure and then retrieves the prior messages in the procedure, including \texttt{CreateSessionRequest} and \texttt{CreateSessionResponse}.

\vspace{2pt}\noindent\textbf{Mapping specification messages to code}. 
After identifying the messages for a candidate, the VA maps them to their code-level handlers by searching the codebase for the message names.
For example, given a candidate in the \texttt{ModifyBearerRequest} handler, the VA identifies two prior messages in the same procedure, \texttt{CreateSessionRequest} and \texttt{CreateSessionResponse}, and then locates their implementations by searching for these names in the codebase; as a result, the S11 \texttt{CreateSessionRequest} handling path and the S5c \texttt{CreateSessionResponse} handling path are discovered, expanding the analysis context from the flagged handler to the full procedure.

\begin{figure}[t]
  \centering
  \includegraphics[width=\linewidth]{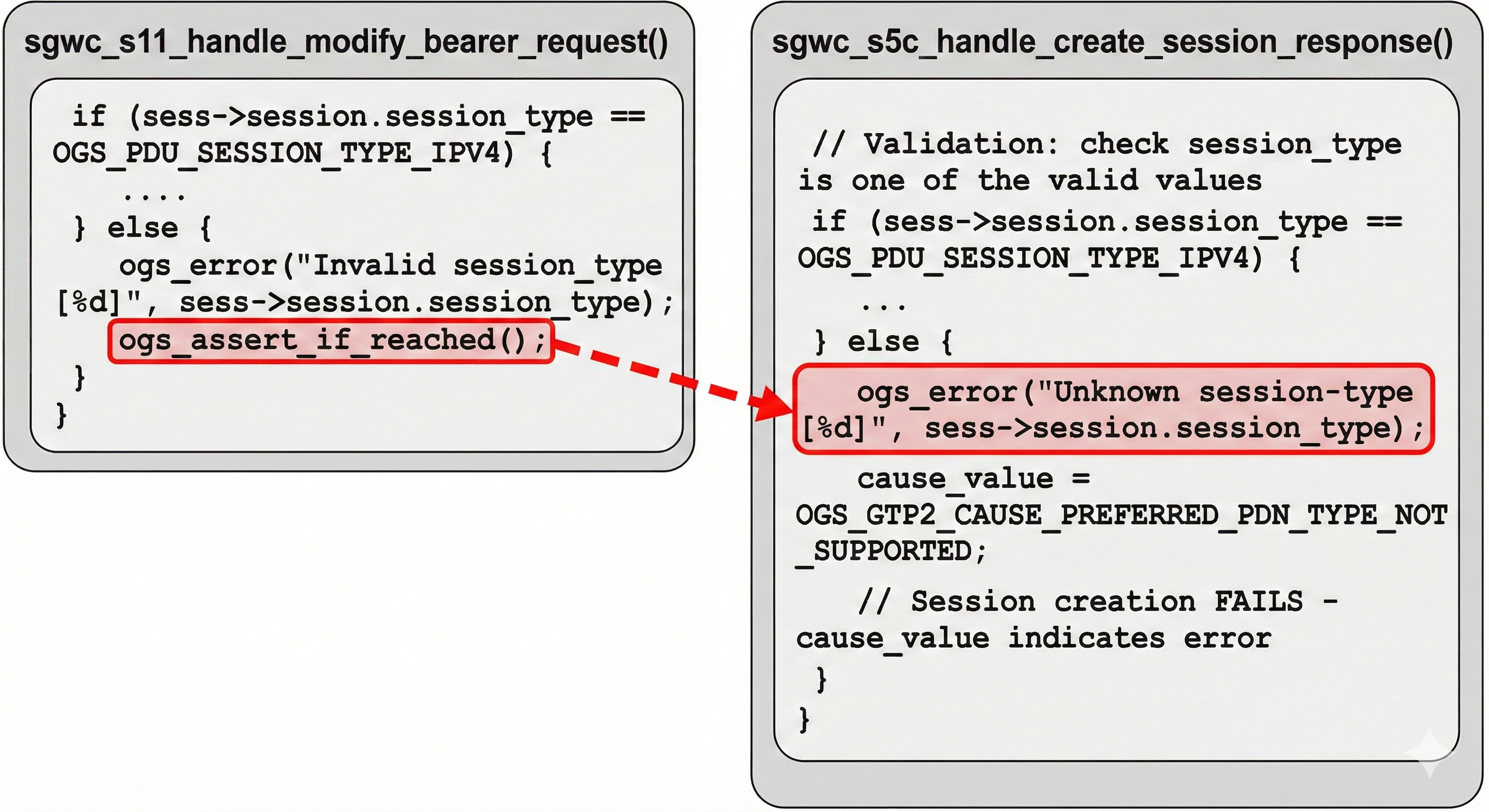}
  \caption{Example of locating an implicit trust violation from a real-world GitHub security issue. The issue description and discussion are used to extract missing-validation assumptions, which are then mapped to the prerequisite messages in the corresponding 3GPP procedure and to the code-level handlers that the VA then reevaluates.}
  \label{fig:crosscheck-case}
\end{figure}

\vspace{2pt}\noindent\textbf{Reevaluating iTrue candidates in expanded context}. 
With the context expanded, the VA reevaluates whether the candidate vulnerability is still exploitable by examining the located handler code for these messages, looking for the relevant validation operations.
For example, consider a candidate that accesses the type field in the PAA IE, where the semantic check on this IE is absent in the handler of \texttt{ModifyBearerRequest} that uses the IE; after expanding the context, however, the VA discovers that this IE is actually a reference to the IE parsed in a prior message handled by \texttt{sgwc\_s5c\_handle\_create\_session\_response()}, which performs both syntactic and semantic checks on the message. This example is illustrated in Figure~\ref{fig:crosscheck-case}.

\subsection{Exploitation Agent}
\label{sec: exploitation agent}
Even after reevaluation, some false positives may still remain, because static reasoning alone cannot guarantee that a candidate is (i) triggerable under a realistic protocol state, (ii) reproducible in an actual deployment, or (iii) genuinely security-impacting. Therefore, \textit{iFinder} invokes the Exploitation Agent (EA) to automatically generate a PoC for each iTrue candidate. More specifically, the EA follows a three-step workflow. First, it derives an explicit attack vector and instantiates protocol-compliant messages using the message schemas extracted during preprocessing, and then generates the PoC strictly from these intermediate artifacts. Second, it performs pre-execution checks (e.g., compilability and consistency with the attack vector) to catch malformed or drifting PoCs. Finally, when a well-formed PoC still fails due to implicit runtime constraints, the EA invokes the \textit{feedback-aware refinement} loop: it analyzes runtime logs, updates the PoC to satisfy missing constraints, and re-executes it until the flaw is triggered or the LLM testing budget is exhausted.

\vspace{2pt}\noindent\textbf{Deriving attack vector and constructing protocol messages}. 
Given an iTrue candidate from the VA (which specifies the vulnerable IE usage site and triggering message type), the EA first derives a message-level attack vector by determining the specific IE field(s) to manipulate and the required content to trigger the target vulnerability. It then constructs the corresponding protocol message(s) according to the schemas, producing well-formed message structures with the attack payloads for PoC generation.
For example, the VA validates a candidate flagging a heap buffer overflow risk introduced by the vulnerable \texttt{memcpy} at \texttt{s5c-handler.c:150}, which copies the \texttt{PAA} IE payload into the buffer \texttt{sess->paa} using the attacker-controlled length field \texttt{paa->len} without bounds checking; from these artifacts, the EA derives the attack vector: sending a \texttt{CreateSessionResponse} containing a malformed \texttt{PAA} IE, where the length field claims 200 bytes while the buffer \texttt{sess->paa} can only hold $\sim$21 bytes, causing an out-of-bounds write; based on the message schema for \texttt{CreateSessionResponse}, the EA populates the required mandatory header fields (e.g., the TEID and sequence number) and embeds the malformed \texttt{PAA} IE at the correct location, constructing a message instance to drive PoC generation.

\vspace{2pt}\noindent\textbf{Generating PoC and performing pre-execution checks}. 
After deriving the attack vector and protocol-compliant messages, the EA generates an initial PoC that instantiates these messages accordingly. Before execution, the EA performs lightweight, tool-assisted pre-execution checks, including (i) syntax checks by compiling the PoC and (ii) consistency checks to ensure the PoC’s message types, field values, and message sequence align with the attack vector.

\vspace{2pt}\noindent\textbf{Executing in a testbed and refining with feedback}. 
The EA then executes the generated PoC by automatically invoking a containerized testbed and analyzing runtime logs to determine whether the vulnerability is triggered. If the attack succeeds (e.g., a crash or an observable state change occurs), the EA confirms the presence of the iTrue vulnerability; otherwise, it diagnoses the failure based on the logs, refines the PoC accordingly, and re-executes it. This execute--analyze--refine loop continues until the vulnerability is triggered or the testing budget is exhausted. For example, an initial PoC may send a malformed \texttt{CreateSessionResponse} with a hardcoded TEID and sequence number. The first execution does not trigger a crash because SGW-C rejects the response: the message cannot be linked to any pending transaction (``unknown TEID''). Based on this feedback, the EA updates the PoC to extract the TEID from \texttt{CreateSessionRequest} and reuse the corresponding sequence number when crafting the response. With this refinement, SGW-C links the response with its pending transaction and processes the \texttt{CreateSessionResponse}, which leads to a crash, confirming the existence of the iTrue.

\section{Evaluation}
\label{sec:evaluation}
We evaluated iFinder on the ground-truth dataset with the known CN iTrues collected from GitHub security issues, and focused on the following research questions:

\noindent\textbf{RQ1} \textit{How effective is iFinder at discovering (known) iTrue vulnerabilities?}

\noindent\textbf{RQ2} \textit{How much does the code-specification cross-checking reduce false positives?}

\noindent\textbf{RQ3} \textit{To what extent does feedback-aware refinement improve PoC generation success?}

\noindent\textbf{RQ4} \textit{What are the time and token costs of running iFinder?}

RQ1 provides an overview of iFinder's effectiveness in detecting iTrue vulnerabilities. RQ2 and RQ3 evaluate the effectiveness of our two main proposed techniques, code-specification cross-checking and feedback-aware refinement, in mitigating LLM hallucinations. RQ4 further examines the efficiency of iFinder.

\subsection{Experimental Setup}

\noindent\textbf{Ground truth dataset.}
As discussed in Section~\ref{sec:preprocessing}, these known iTrue vulnerabilities are classified into six patterns with the following distribution: \textit{Malformed Field} (11), \textit{Absent Field} (1), \textit{Invalid Value} (6), \textit{Invalid State} (1), \textit{Invalid Reference} (1), and \textit{Resource Exhaustion} (2). The distribution is heavy-tailed, with a small number of patterns accounting for most instances.

\noindent\textbf{Tested CN implementations.}
We evaluated iFinder on two open-source CN implementations, Open5GS and free5GC. Since each ground-truth vulnerability exists only in a specific range of versions, we selected multiple version snapshots to cover the affected releases referenced by the corresponding issues. Specifically, we used Open5GS v2.7.5, v2.7.2, and v2.4.14, as well as free5GC v3.3.0 and v2.0.2. This setup ensures that each vulnerability is evaluated on a version in which it is known to exist. 

\noindent\textbf{Evaluation metrics.}
For vulnerability detection, we report true positives (TP), false positives (FP), and false negatives (FN), along with precision, recall, and F1.
A report is counted as a true positive (TP) if it correctly identifies a ground-truth vulnerability; similarly, we define false positives (FPs) and false negatives (FNs).
For PoC generation, we consider a PoC successful if it triggers the expected behavior.

\noindent\textbf{System Implementation}.
We implement iFinder's multi-agent pipeline using the Claude Agent SDK. All methods, including the prompt-only baseline, use Claude Opus 4.5 as the underlying LLM. We ran all methods with the default configurations (e.g., decoding parameters and context limits) and capped the number of refinement iterations per vulnerability to five. We measured runtime as wall-clock time and reported token usage as the total input and output tokens consumed by all agents (including refinement), which we further converted to cost using the pricing of the provider.
All experiments are repeated five times, and we report the average results.

\subsection{RQ1: Detection Effectiveness}

To answer RQ1, we evaluated iFinder's ability to identify the 22 known vulnerabilities and compared its performance against a prompt-only baseline.
Notably, this baseline is not a pure LLM setup; it is fed with the patterns we summarized from GitHub issues. We adopted this enriched baseline because directly asking the LLM to discover vulnerabilities yielded severe hallucinations and no usable results.
We report TP and FN to compute recall, which measures how well the approach identifies the known vulnerabilities. In addition, we report the number of newly discovered vulnerabilities and FP among all findings, from which we compute precision.
The results are summarized in Table~\ref{tab:effectiveness}. 
Overall, iFinder achieved an F1 score of $71.429\%$, which is more than twice that of the baseline.
More specifically, iFinder identified $68.182\%$ of the ground-truth vulnerabilities, which is roughly twice as high as the baseline.
To understand the remaining gap, we manually reviewed all seven missed cases and attributed them to two root causes.
First, \textit{pattern coverage gaps}: four cases involve corner-case validation logic not captured by the generalized detection patterns distilled from cluster-level root causes. 
Second, \textit{incomplete context construction}: three cases lie on long cross-module execution paths whose caller chains the DA cannot reliably reconstruct, so the missing check cannot be confirmed and therefore the candidate was not flagged.
Notably, while the baseline performed substantially worse, it did not completely lose utility, achieving $28.205\%$ precision. 
This is because the baseline incorporates the patterns we summarized, which also highlights the utility of these patterns in guiding LLM-based vulnerability discovery.
In contrast, more than $2/3$ of the reported vulnerabilities were confirmed to be real.
In a later section, we further discuss how our proposed code-specification cross-checking improves precision in detail.

\begin{table}[!t]
  \centering
  \caption{Detection performance on the 22 ground-truth vulnerabilities.}
  \resizebox{0.48\textwidth}{!}{
    \begin{tabular}{lccccccc}
    \toprule
    Configuration & TP & FP & FN & New & Precision & Recall & F1\\
    \midrule
    \textbf{Prompt-only} & 8 & 56 & 14 & 14 & 28.205\% & 36.364\% & 31.769\% \\
    \textbf{iFinder} & 15 & 12 & 7 & 21 & 75.000\% & 68.182\% & 71.429\% \\
    \bottomrule
    \end{tabular}
  }
  \label{tab:effectiveness}
\end{table}

\subsection{RQ2 + RQ3: Ablation Study}
\label{sec:ablation}

To answer RQ2 and RQ3, we conducted an ablation study to evaluate the effectiveness of our two proposed primitives: code-specification cross-checking, performed by the VA, and feedback-aware refinement, performed by the EA. 
Comparing DA, DA+VA, and DA+VA+EA (Table~\ref{tab:ablation}), we found that code-specification cross-checking reduced the FP from 62 to 19, improving the precision from 36.735\% to 65.455\%.
To assess feedback-aware refinement, we provided the 22 ground-truth vulnerabilities to the EA and to the prompt-only baseline (i.e., directly prompting the LLM) and measured how well each generated valid PoCs.
The EA produced valid PoCs for 19 of the 22 ground-truth vulnerabilities, whereas the prompt-only baseline succeeded on only 8, demonstrating the effectiveness of feedback-aware refinement in enabling reliable PoC generation.
The EA further reduced FP from 19 to 12 by discarding candidates whose PoCs fail to trigger the intended flaw, yielding the full-pipeline precision of 75\%.
Still 12 FPs remained because a generated PoC can trigger a real yet unintended vulnerability located prior to the target flaw on the execution path. 
The EA only checks whether a PoC produces the expected runtime effect, which in the case of an FP, is always a crash. 
It does not verify whether the crash site corresponds to the intended vulnerability candidate.
We identify these cases by manual inspection: for each PoC the EA reports as successful, we compare the \textit{observed crash site} from the runtime logs against the \textit{expected crash site}, i.e., the dangerous operation and its handler path localized by the DA and the VA. 
If the observed crash site does not match the expected crash site, the PoC is believed to have triggered an unintended vulnerability, so we count the report as an FP.

\begin{table}[!t]
  \centering
  \caption{Ablation study.}
  \resizebox{0.48\textwidth}{!}{
    \begin{tabular}{lccccccc}
    \toprule
    Configuration & TP & FP & FN & New & Precision & Recall & F1\\
    \midrule
    \textbf{DA}    & 15 & 62 & 7 & 21 & 36.735\% & 68.182\% & 47.745\% \\
    \textbf{DA+VA} & 15 & 19 & 7 & 21 & 65.455\% & 68.182\% & 66.790\% \\
    \textbf{DA+VA+EA} & 15 & 12 & 7 & 21 & 75.000\% & 68.182\% & 71.429\% \\
    \bottomrule
    \end{tabular}
  }
  \label{tab:ablation}
\end{table}

\subsection{RQ4: Overhead Analysis}

To answer RQ4, we measured the runtime and token usage of each pipeline stage; the results are summarized in Table~\ref{tab:performance}.
The Discovery Agent accounted for the majority of the runtime (39.436\%), as it scans the codebase to locate message handlers and performs LLM-based backward analysis before producing vulnerability candidates. Vetting was comparatively lightweight (32.394\%) because it operates only on DA vulnerability candidates and retrieves minimal protocol-aligned context. Exploitation time varies with the number of refinement iterations (up to 5).

\begin{table}[ht]
  \centering
  \caption{Time and token breakdown per target implementation.}
  \small
  \begin{tabular}{@{}lcc@{}}
    \toprule
    \textbf{Stage} & \textbf{Avg. Time (s)\,$\downarrow$} & \textbf{Tokens\,$\downarrow$} \\
    \midrule
    Discovery     & 280  & 2.1M \\
    Vetting       & 230  & 3.3M \\
    Exploitation  & 200  & 4.4M \\
    \midrule
    \textbf{Total} & \textbf{710} & \textbf{9.8M} \\
    \bottomrule
  \end{tabular}
  \label{tab:performance}
\end{table}

\section{Measurement on New CN iTrues}
\label{sec:measurement}

We further ran \textit{iFinder} on seven widely used open-source CN implementations, including Open5GS 5G, free5GC, OAI 5G, SD-Core, eUPF, Open5GS LTE, and OAI LTE, covering both 4G and 5G systems and in three different languages (C, C++, and Go). Our goal is to assess whether the discovered vulnerabilities are isolated implementation bugs or instead indicate systemic weaknesses in modern core networks.

\subsection{Discoveries}

\begin{table}[!t]
  \centering
  \caption{Vulnerability discovery results by implementation.}
  \resizebox{0.49\textwidth}{!}{
    \begin{tabular}{lcccc}
    \toprule
    Target & Language &\#Discovered & \#Confirmed & \#CVE \\
    \midrule
    Open5GS 5G &C &10 & 9 & 9 \\
    free5GC &Go &14 & 14 & 12 \\
    OAI 5G &C++ &11 & 11 & 11 \\
    SD-Core &Go &7 & 7 & 7 \\
    eUPF &Go &5 & 5 & 5 \\
    \midrule
    Open5GS LTE &C & 30 & 30 & 30 \\
    OAI LTE &C++ & 7 & 7 & 7 \\
    \midrule
    \textbf{Total} & &\textbf{84} & \textbf{83} & \textbf{81} \\
    \bottomrule
    \end{tabular}
  }
  \label{tab:overall_results}
\end{table}

\noindent\textbf{Landscape}. Across the seven evaluated implementations, \textit{iFinder} reported 84 previously unknown vulnerabilities spanning two protocols, GTP-C and PFCP (Table~\ref{tab:overall_results}). \textit{At the time of writing, 83 findings have been confirmed by developers, and 81 have been assigned CVE identifiers}, spanning all seven implementations. 
While we do not claim to have exhaustively captured all iTrue vulnerabilities in these codebases, the pervasiveness of the flaws, and their recurrence across multiple systems (implemented in different programming languages) and across different protocols, strongly suggests that iTrues are not isolated implementation bugs but rather systemic weaknesses in carrier core networks, rooted in implicit trust in internal peers. As discussed earlier, this assumption is increasingly invalidated as CNs migrate to less protected cloud environments.

Of particular interest, we observe that certain iTrue flaws in 5G systems appear to be inherited from their 4G counterparts. This suggests that the new risk can arise from a failure to adapt legacy designs and assumptions to a deployment environment that is no longer equally trusted. Note that many production CNs today support both 4G and 5G simultaneously, so risks originating in 4G components remain a serious security concern.

What is most surprising is the diversity of vulnerabilities that nonetheless fall under the same high-level category. For example, in free5GC, repeatedly sending PFCP \texttt{Session Establishment Request} messages can drive the UPF into a crash by exhausting internal resources.
In Open5GS, repeatedly sending GTPv2-C \texttt{Create Session Request} messages can similarly cause the SGW-C to terminate by depleting multiple internal resource pools, including UE contexts, transaction timers, and bearer-related structures. Despite differences in the targeted resources, protocols, and implementation languages, both cases can be attributed to the same root cause: handlers implicitly trust that internal requestors will behave prudently and will not intentionally deplete finite resources. This optimism often leads to fragile exception handling that is incomplete or inconsistently applied. As a result, resource exhaustion is more likely to surface as a process crash than as controlled rejection or graceful degradation. Notably, despite their diversity, both flaws are captured by the same detection pattern.

\begin{table}[!t]
  \centering
  \caption{Cross-protocol distribution of vulnerability patterns.}
  \resizebox{0.38\textwidth}{!}{
    \begin{tabular}{lccc}
    \toprule
    Pattern & GTP-C & PFCP & Total \\
    \midrule
    Malformed Field & 11 & 3 & 14 \\
    Absent Field & 6 & 25 & 31 \\
    Invalid Value & 4 & 8 & 12 \\
    Invalid State & 7 & 1 & 8 \\
    Invalid Reference & 2 & 3 & 5 \\
    Resource Exhaustion  & 7 & 7 & 14 \\
    \midrule
    \textbf{Total} & \textbf{37} & \textbf{47} & \textbf{84} \\
    \bottomrule
    \end{tabular}
  }
  \label{tab:pattern_distribution}
\end{table}

\vspace{2pt}\noindent\textbf{Cross-protocol pattern distribution}. To characterize the discovered vulnerabilities, we analyzed their distribution across pattern families, as summarized in Table~\ref{tab:pattern_distribution}. Notably, every pattern family recurs in both GTP-C and PFCP, suggesting that many vulnerabilities arise from shared implementation assumptions rather than protocol-specific semantics.
Among all patterns, \textit{Absent Field} is the most common, accounting for 31 of the 84 vulnerabilities, with a substantial fraction occurring in PFCP implementations (25/31). This result is consistent with PFCP handlers implicitly assuming the presence of mandatory information elements and proceeding along nominal control procedures without enforcing completeness checks.
\textit{Malformed Field} is another prevalent pattern, reflecting insufficient validation of length fields and structural counts in complex message formats. Compared with PFCP, GTP-C exhibits more \textit{Malformed Field} vulnerabilities, likely because GTP-C messages rely on a richer and more diverse IE ecosystem, with extensive use of grouped IEs and nested length fields. This complexity makes GTP-C parsers more susceptible to desynchronization, which can in turn trigger memory-safety issues or logic errors.
In addition, \textit{Invalid Value} and \textit{Invalid State} show a skewed distribution: \textit{Invalid Value} is more prevalent in PFCP, whereas \textit{Invalid State} occurs more often in GTP-C. This asymmetry may reflect the more complex state machines managed by GTP-C implementations, which are more susceptible to inconsistencies under unexpected message sequences. By comparison, \textit{Invalid Reference} and \textit{Resource Exhaustion} are more evenly distributed across the two protocols, likely because both protocols manage broadly comparable categories of resources and therefore exhibit similar exposure to \textit{Invalid Reference} and \textit{Resource Exhaustion} vulnerabilities.

Interestingly, although PFCP is a newer protocol originally introduced for CUPS and later widely adopted in 5G architectures, it exhibits more vulnerabilities than GTP-C in our study (47 versus 37). One plausible explanation is that PFCP handlers are often implemented as internal resource-management interfaces, where messages are more likely to be assumed well-formed and thus receive less rigorous validation.
Overall, the cross-protocol recurrence of syntax-, semantics-, and resource-related pattern families suggests that implicit-trust assumptions can persist at the implementation level across protocol generations, leading to accumulated gaps in validation logic rather than isolated programming errors.

\begin{table}[!t]
  \centering
  \caption{Vulnerability breakdown by CN components.}
  \resizebox{0.26\textwidth}{!}{
    \begin{tabular}{lcccc}
    \toprule
    \multirow{2}{*}{Component} & \multicolumn{2}{c}{Protocol} & \multirow{2}{*}{Total} \\
    \cmidrule(r){2-3}
     & GTP-C & PFCP & \\
    \midrule
    \multicolumn{4}{l}{\textit{4G Core}} \\
    \quad MME & 2 & N/A & 2 \\
    \quad SGW & 32 & 1 & 33 \\
    \quad PGW & 3 & 0 & 3 \\
    \midrule
    \multicolumn{4}{l}{\textit{5G Core}} \\
    \quad SMF & N/A & 9 & 9 \\
    \quad UPF & N/A & 37 & 37 \\
    \midrule
    \textbf{Total} & \textbf{37} & \textbf{47} & \textbf{84} \\
    \bottomrule
    \end{tabular}
  }
  \label{tab:nf_breakdown}
\end{table}

\vspace{2pt}\noindent\textbf{Component-level breakdowns}. To understand where these implicit-trust assumptions most frequently surface in practice, we further examined how vulnerabilities are distributed across core network components, with results summarized in Table~\ref{tab:nf_breakdown}.
The results also revealed a skewed distribution. In 4G systems, the Serving Gateway (SGW) accounted for the majority of GTP-C vulnerabilities (32/37). In 5G systems, the User Plane Function (UPF) accounted for the majority of PFCP vulnerabilities (37/47).
Such observed concentration is because SGW/UPF components maintain long-lived protocol state across multi-step procedures, manage scarce and performance-critical resources, and mediate interactions among multiple peers while reconciling control-plane decisions with user-plane resource allocation. Such cross-plane coupling increases the complexity of state transitions and error handling, which can make missing validation and incomplete recovery paths more likely to surface.
Moreover, the discovered flaws span both 4G and 5G, and this cross-generation coverage yet again supports the view that iTrue reflects a systemic weakness rather than isolated bugs in individual codebases: these components sit at the intersection of stateful control logic and resource management.

\subsection{Case Studies}
\label{sec: case studies}
We present two case studies to demonstrate the security impact of the newly discovered iTrue vulnerabilities.

\begin{figure}[t]
  \centering
  \includegraphics[width=0.85\linewidth]{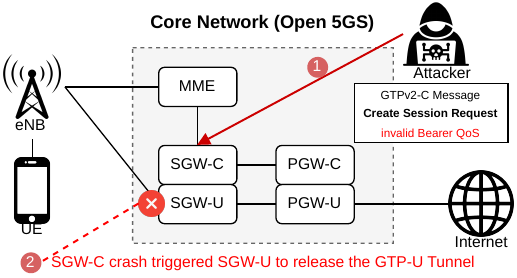}
  \caption{Example attack exploiting malformed GTPv2-C messages to trigger denial-of-service in a core network component by abusing missing validation checks.}
  \label{fig: dos}
\end{figure}

\noindent\textbf{Denial of service}.
\textit{iFinder} reported an iTrue vulnerability in SGW-C caused by missing syntactic validation of the Bearer Quality of Service (QoS) Information Element (IE) when parsing GTPv2-C \textit{Create Session Request} messages. Specifically, a \textit{Create Session Request} that carries a Bearer QoS IE with an invalid fixed-length encoding can trigger an assertion, causing SGW-C to crash and resulting in a DoS.
As stated in our threat model, we consider a remote adversary with no direct control over cellular CN components, but who can send arbitrary GTPv2-C messages to trigger the vulnerability. In our study, we successfully executed an end-to-end DoS attack against Open5GS LTE in our testbed. As shown in Figure~\ref{fig: dos}, the attack proceeds as follows: (1) the attacker sends a GTPv2-C \textit{Create Session Request} containing a Bearer QoS IE with an invalid fixed-length encoding; (2) when SGW-C parses the malformed IE, it crashes immediately, triggering SGW-U to release the SGW-U-eNB GTP-U tunnels.

\vspace{2pt}\noindent\textbf{Session hijacking}. 
\textit{iFinder} detected an iTrue vulnerability in the UPF caused by missing uniqueness validation of Packet Detection Rule (PDR) IDs when processing PFCP \textit{Session Modification Request} messages. Specifically, by injecting a duplicate PDR ID with a lower \texttt{Precedence} value (i.e., higher priority), an attacker can overrule the legitimate rule. This logic flaw causes the UPF to match the malicious PDR first during packet processing, resulting in session hijacking.

\begin{figure}[t]
  \centering
  \includegraphics[width=0.95\linewidth]{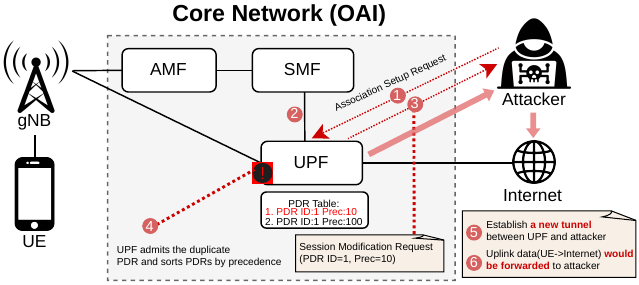}
  \caption{Example attack exploiting duplicate PDR IDs in PFCP \textit{Session Modification Request} messages to trigger session hijacking in UPF by abusing missing uniqueness validation.}
  \label{fig: session hijacking}
\end{figure}

In line with the threat model, we assume an Internet attacker with no direct control over core-network components who, through a cloud-native misconfiguration that exposes the N4 interface, can send arbitrary PFCP messages to the UPF to trigger a vulnerability.
We demonstrated an end-to-end session-hijacking attack on OAI 5G in our testbed. 
As shown in Figure~\ref{fig: session hijacking}, the attack proceeds as follows: (1) the attacker sends a PFCP \textit{Association Setup Request} to the UPF; (2) the victim UE initiates an attach, triggering the SMF to send a PFCP \textit{Session Establishment Request} to the UPF; (3) the attacker then issues a PFCP \textit{Session Modification Request} that reuses the victim's PDR ID with a lower \texttt{Precedence} value (higher priority) and binds it to a malicious Forwarding Action Rule (FAR); (4) the UPF admits the duplicate PDR and sorts PDRs by precedence, placing the malicious rule ahead of the legitimate one; and (5) during packet processing, the UPF matches the malicious PDR first, and establishes a new forwarding tunnel between UPF and attacker; and (6) as a result, the victim's uplink traffic is forwarded to the attacker rather than to the Internet.  
Beyond the OAI testbed, we further validated this flaw on two commercial 5G core networks. In collaboration with a partner vendor, we reproduced the attack in the vendor's lab environment under default configurations. In both cases, the same root cause (missing uniqueness validation of PDR IDs) was present, and the attack exhibited the same session-hijacking behavior observed on OAI. Specifically, the duplicate PDR with a lower \texttt{Precedence} value (higher priority) overruled the legitimate rule for forwarding the victim's uplink traffic to the Internet, redirecting it to the attacker instead. One vendor has fixed the issue, and a CVE has been assigned (CVE-2026-8233); the other, a major 5G carrier, is still in the remediation process.

\section{Discussion}

\noindent\textbf{Vulnerabilities in CNs}. Our research brings to light the pervasiveness and severity of iTrue vulnerabilities, which we attribute to a systemic specification-to-implementation gap. Requirements for protocol-syntax validation, protocol-semantics validation, and resource-availability checks are dispersed across 3GPP specifications, increasing the likelihood of incomplete enforcement in implementations and leading to recurring weaknesses such as iTrue. Crucially, these gaps threaten not only availability (e.g., denial of service) but also integrity (e.g., session hijacking), as demonstrated in our research, indicating that iTrue flaws are not isolated implementation bugs but systematic concerns.

Mitigating iTrue risks requires coordinated action across specifications, implementation, and deployment. At the specification level, 3GPP should update its threat model to reflect cloud-native deployments in which internal interfaces may be exposed, and should mandate mutual authentication on internal interfaces. At the implementation level, network-function developers should enforce strict syntactic and semantic validation at protocol boundaries, maintain uniqueness invariants for session and rule identifiers, and adopt bounded resource allocation with graceful rejection rather than process-ending assertions. At the deployment level, operators should adopt a zero-trust posture by removing unintended reachability, enforcing net-segmentation and least-privilege access between components, and continuously auditing cloud configurations to detect misconfigurations before they become exploitable.

We hope that the findings of this study will help catalyze such efforts, contributing to defense-in-depth protections for future CNs in particular, and carrier networks more broadly.

\vspace{2pt}\noindent\textbf{Limitations of iFinder}. 
\textit{iFinder} does not guarantee complete coverage of all iTrue vulnerabilities. The Discovery Agent is bounded by our seed pattern set, and vulnerabilities that do not match these patterns may be missed. Moreover, although our pipeline leverages procedure context and message schemas to reduce false positives, the LLM may still make incorrect feasibility judgments when exploitability hinges on subtle specification semantics, implementation-specific state transitions, or implicit invariants that are not captured by message structure alone. Future work should explore additional implicit assumptions underlying CNs and incorporate deeper, system-level analysis to further reduce false positives and broaden coverage.

\vspace{2pt}\noindent\textbf{Agent-based security analysis on carrier networks}. 
Our results demonstrate that multi-agent systems can automate vulnerability discovery across multiple protocols, achieving a level of scalability and accuracy that is difficult to attain with manual auditing or conventional, less intelligent tools. Of particular importance is the agentic system’s ability to perform multimodal analysis, jointly reasoning over code, specifications, and other artifacts, and to do so smoothly across different programming languages and protocols.

These capabilities are especially valuable for carrier-network security analysis, where the ecosystem is complex and heterogeneous, and where both the codebases and the supporting documentation are massive. Looking ahead, we expect agentic techniques to see broader adoption and tighter integration into pipelines for both specification development and system implementation, ultimately improving the security posture of this critical information infrastructure.

\section{Related Work}
\vspace{2pt}\noindent\textbf{Cellular Network Security.}
A large body of research has examined the security of cellular networks across different protocol layers and interfaces. At the UE and RAN side, prior work has investigated baseband vulnerabilities~\cite{kim2021basespec, tu2024logic, maier2020basesafe}, lower-layer attacks~\cite{rupprecht2019breaking, yang2019hiding}, eavesdropping~\cite{rupprecht2020call}, privacy violations~\cite{hussain2019privacy}, and protocol compliance testing~\cite{kim2019touching, hussain2021noncompliance, park2022doltest, khandker2024astra, yang2024oranalyst}.
However, these efforts predominantly target the air interface between UE and RAN or the NAS/RRC signaling between UE and core, leaving the \textit{internal interfaces} of the core network relatively underexplored.
Recent work has started to examine core-network security from specific angles:
formal verification of access-control mechanisms~\cite{akon2023formal, akon2025control},
provenance-based attack detection~\cite{pacherkar2024prov5gc},
context-integrity testing of the LTE core~\cite{son2025citesting},
and GTP-U tunnel abuse that demonstrates external attackers can reach internal CN interfaces~\cite{zhang2025invade, shaik2025uncovering}.
Motivated by the growing reachability of internal interfaces in cloud-native deployments, our work takes a complementary direction: we systematically study implementation vulnerabilities across PFCP and GTP-C protocols and characterize \textit{iTrue} as a recurring type of security flaw, where core-network components blindly trust internal messages and skip critical validations, leading to severe consequences including DoS and session hijacking.

\vspace{2pt}\noindent\textbf{Automated Vulnerability Discovery in Cellular Networks.}
Fuzzing has been applied to cellular implementations from UE-RAN side testing~\cite{kim2019touching, park2022doltest, garbelini2022towards} to core-network fuzzing~\cite{bennett2024ransacked,dong2025corecrisis, sun20255gc, yang2024feedback}, but these approaches predominantly uncover memory-safety bugs and crash-inducing issues, while many high-impact core-network failures are logic vulnerabilities that do not manifest as crashes.
Specification-based analysis mines vulnerabilities from 3GPP documents through NLP~\cite{chen2021bookworm}, change-request analysis~\cite{chen2022seeing}, automated reasoning~\cite{chen2023sherlock}, undefined-behavior studies~\cite{klischies2023instructions}, and protocol state-machine synthesis~\cite{alishtiaq2024hermes, rahman2024cellularlint, xie2025cellsecinspector}; 
however, these methods target design flaws at the specification level and cannot automatically validate whether identified issues are exploitable in real implementations.
In contrast, iFinder distills recurring vulnerability patterns from previously discovered core network flaws and uses them to guide an LLM-driven multi-agent system, enabling efficient discovery of both memory-safety and deep logic vulnerabilities while closing the loop with code-specification cross-checking and runtime-verified PoC generation.

\vspace{2pt}\noindent\textbf{Multi-Agent for Vulnerability Discovery.}
LLMs have recently been applied to automate vulnerability discovery through agentic and multi-agent pipelines. 
Previous works have used LLMs to enhance existing static-analysis tools, e.g., improving the efficiency of CodeQL~\cite{wang2026cqllm} or customizing CodeQL for JavaScript vulnerability detection~\cite{ghebremichael2026multi}.
Others build general-purpose multi-agent pipelines to discover bugs in general programs~\cite{wu2026mulvul, liu2026synthesizing}.
However, these systems target general-purpose code and are not tailored to stateful protocols, so they are ineffective for 5G core networks and only report vulnerability candidates without confirming exploitability.
In contrast, \textit{iFinder} employs a pattern-guided multi-agent system that performs code-specification cross-checking to recover procedure context and iteratively refines PoCs with feedback, enabling it to detect iTrues in 5G core network implementations and confirm their exploitability.

\vspace{2pt}\noindent\textbf{Input Validation and Specification-Driven Testing.}
Some iTrues are related to input-validation vulnerabilities, a broad class that includes XSS~\cite{wassermann2008static}, SQL injection~\cite{al2023sqirl}, and, more recently, prompt injection~\cite{greshake2023not}, each with distinct root causes and mitigation strategies.
However, many other iTrues do not fall under input validation: they are triggered by well-formed inputs that arrive in incorrect system states. Because PFCP and GTP-C are long-lived, stateful protocols, inconsistencies between internal states and otherwise valid inputs can introduce security risks.
For example, a duplicate PDR ID may be syntactically valid yet still cause session hijacking. This distinction differentiates many iTrues from input-validation vulnerabilities.
Moreover, although iTrues are not fundamentally new, their detection cannot be effectively addressed by existing techniques. Specification-driven approaches, including model-based testing~\cite{utting2012taxonomy}, primarily target design flaws, while stateful protocol testing~\cite{ba2022stateful} and stateful network fuzzing~\cite{pham2020aflnet} often rely on crash-based signals and therefore struggle to capture logic flaws that do not manifest as crashes.

\section{Conclusion}
As cloud-native development erodes the physical isolation traditionally afforded to cellular CNs, we use \textit{implicit trust errors} (iTrue) to capture a recurring type of security flaw. We designed \textit{iFinder}, an LLM-driven multi-agent system, and applied it to discover new iTrues in CN implementations. Running iFinder on seven prominent open-source CN implementations, we discovered 84 previously unknown vulnerabilities. Among them, 83 have been confirmed with 81 CVE assignments already awarded. Importantly, a session-hijacking flaw has already been confirmed on real-world commercial 5G core networks. Our findings highlight the pervasiveness of iTrues across the cellular CNs and the urgent need for elevated protection within the original trust domains. We plan to release iFinder as a public service to help enhance the security of cellular core networks.

\section*{Acknowledgments}
We thank the shepherd, all anonymous reviewers for their valuable comments. We also thank Dr.~Dongfang Zhao for his helpful discussions and insights on leveraging large language models for vulnerability detection. This research is supported by the National Research Foundation, Singapore, and the Cyber Security Agency of Singapore under the National Cybersecurity R\&D Programme and the CyberSG R\&D Programme Office (Award CRPO-GC3-NTU-001) and (Award CRPO-GC6-NTU-001), and NTU startup grant (025559-00001). Any opinions, findings, conclusions, or recommendations expressed in these materials are those of the author(s) and do not reflect the views of the National Research Foundation, Singapore, the Cyber Security Agency of Singapore, or the CyberSG R\&D Programme Office.

\section*{Open Science}
We make the artifacts required to evaluate our contributions available at this link: \url{https://zenodo.org/records/20534406}.
These artifacts include the implementation of the iFinder framework, the curated dataset of publicly reported iTrue vulnerabilities used as ground truth, and scripts and configuration files necessary to reproduce the main experimental results in controlled environments. Beyond the reproducibility artifacts, we also maintain a public website at \url{https://linziyuu.github.io/iFinder-Website/} that compiles all 84 findings.
To prevent misuse prior to widespread patch adoption, we do not release exploit-grade proof-of-concept code for newly discovered vulnerabilities. The released artifacts are sufficient to reproduce our evaluation and validate the reported findings. 

\section*{Ethical Considerations}
We structure the ethical considerations discussion by linking our stakeholder analysis to the impacts generated during two distinct phases: the \textit{research process} (vulnerability discovery and validation) and the \textit{publication of results} (disclosure and artifact release). We then detail \textit{mitigations} (specifying which stakeholder group each measure protects) and conclude with the \textit{justification} for conducting this research.

\noindent\textbf{Stakeholder Analysis and Process Impact.} Our work involves six primary stakeholder groups. 
(1) \textit{Open-source CN Maintainers}: the developer teams of the seven evaluated implementations (Open5GS 5G, Open5GS LTE, free5GC, OAI 5G, OAI LTE, SD-Core, and eUPF). They rely on the research process to receive clear, reproducible, and responsibly timed vulnerability reports that enable effective patching. 
(2) \textit{Commercial CN Vendors}: two commercial core-network vendors whose products we additionally tested under vendor-sanctioned conditions. They rely on confidential disclosure channels to protect their customers during coordinated remediation. 
(3) \textit{Network Operators and Deployers}: telecom operators, private-5G tenants, and research deployments running the affected stacks; their operational security depends on whether discovered flaws are responsibly disclosed and remediated before public release. 
(4) \textit{End Users}: although no user was involved in our experiments, subscribers are the ultimate stakeholders whose confidentiality, availability, and session integrity are at stake when CN flaws are exploited. 
(5) \textit{3GPP/GSMA}: their specifications encode the implicit-trust assumptions targeted by iFinder, and they rely on research findings to inform future threat models. 
(6) \textit{Researchers}, who rely on methodologically sound and ethically scoped procedures. During the research process, all experiments were conducted in isolated, self-hosted testbeds using open-source implementations or vendor-authorized evaluation units. No production operator network, live subscriber, or real user data was accessed, and no stakeholder experienced availability or privacy impact from our experimentation.

\noindent\textbf{Impact of the Research.} The publication of this work has both positive and negative impacts, affecting the stakeholders above.

\noindent\textit{Positive Impacts.} (1) Strengthening Deployed Infrastructure (Impact on Maintainers, Vendors, Operators \& End Users): iFinder produced 84 previously unknown vulnerability reports, of which 83 have been confirmed and 81 have been assigned CVEs, directly improving the security of widely deployed CN stacks. As a concrete indicator that affected stakeholders view the work as net-beneficial, one of the evaluated projects has proactively requested continued access to iFinder for long-term automated scanning. (2) Informing Standards Bodies (Impact on 3GPP/GSMA): by naming and characterizing iTrue as a recurring specification-to-implementation gap rather than isolated bugs, we equip 3GPP/GSMA, operators, and defenders with a reusable lens to audit their stacks. (3) Reproducibility and Open Science (Impact on Researchers): we release the iFinder framework and the curated ground-truth dataset to allow independent replication, extension, and scrutiny.

\noindent\textit{Negative Impacts.} (1) Misuse of Automated Discovery (Impact on Operators \& End Users): the iFinder pipeline lowers the cost of finding similar flaws in other stateful protocol stacks and could be repurposed by malicious actors if released without safeguards. (2) Facilitating Analogous Attacks (Impact on Operators \& End Users): public characterization of the vulnerabilities may help adversaries search for analogous bugs, especially in deployments that have not yet applied upstream fixes.

\noindent\textbf{Mitigation.} To address the risks outlined above, we have implemented the following measures, each tied to the stakeholder it protects. 
(1) \textit{Coordinated Disclosure for Maintainers \& Operators}: every confirmed vulnerability was disclosed to the respective upstream project or vendor prior to submission. 
As of the response deadline, 81 CVEs have been assigned, and 58 of 83 confirmed vulnerabilities in open-source implementations have been patched; the per-implementation status is summarized in Table~\ref{tab:disclosure-status}.
Vulnerabilities on the two commercial CNs were disclosed directly to the vendors.

One affected commercial 5GC vendor has fixed the issues, and CVE identifiers have been assigned, including session hijacking (CVE-2026-8233) and DoS (CVE-2026-8232). 
The second commercial 5GC vendor, which is a major 5G carrier, is still in the remediation process.
(2) \textit{Reducing Misuse Risk for Operators}: although iFinder includes an Exploitation Agent for validating candidate findings in controlled testbeds, the public artifact will not include a collection of PoCs, and we will not provide PoCs or reproduction details for unpatched vulnerabilities. 
(3) \textit{Protecting Commercial Vendors}: for any vendor still under coordinated remediation, we anonymize vendor identity. 
(4) \textit{Controlled Experimental Scope for End Users}: all experiments were executed in isolated testbeds with no interaction with production networks, commercial operator infrastructure, or real subscribers.

\begin{table}[!t]
  \centering
  \caption{Coordinated-disclosure status per open-source implementation.}
  \label{tab:disclosure-status}
  \resizebox{0.49\textwidth}{!}{
    \begin{tabular}{lcccc}
    \toprule
    Implementation & \#Discovered & \#Confirmed & \#CVE & \#Fixed \\
    \midrule
    Open5GS 5G  & 10 & 9  & 9  & 9  \\
    free5GC     & 14 & 14 & 12 & 12 \\
    OAI 5G      & 11 & 11  & 11  & 0  \\
    SD-Core     & 7  & 7  & 7  & 7  \\
    eUPF        & 5  & 5  & 5  & 0  \\
    \midrule
    Open5GS LTE & 30 & 30 & 30 & 30 \\
    OAI LTE     & 7  & 7  & 7  & 0  \\
    \midrule
    \textbf{Total} & \textbf{84} & \textbf{83} & \textbf{81} & \textbf{58} \\
    \bottomrule
    \end{tabular}
  }
\end{table}

\noindent\textbf{Justification for Research.} This work addresses a critical and systemic security challenge in carrier core networks: as CNs migrate to cloud-native deployments, the long-standing implicit-trust assumption between internal network functions no longer holds, exposing end users and operators to risks that have received far less scrutiny than RAN or UE components. 
Conducting this research is essential because: (1) it leads to the discovery of unknown vulnerabilities in open-source implementations and facilitates the enhancement of their security; (2) characterizing iTrue as a recurring type of security flaw, rather than reporting isolated bugs, enables standards bodies, vendors, and defenders to generalize mitigations beyond the specific cases we found; 
(3) releasing the iFinder framework fosters transparency and community-driven improvements, reducing the risk of opaque or unilateral vulnerability discovery. 
Under these constraints, we believe the benefits of systematically identifying and mitigating recurring implicit-trust failures in critical network infrastructure outweigh the residual risks.

\bibliographystyle{unsrt}
\normalem
\bibliography{reference}

\end{document}